\documentstyle[12pt,aaspp4]{article}

\begin{document}

\title{Cosmology in a String-Dominated Universe}
\author{David Spergel}
\affil{Princeton University Observatory, Princeton, NJ 08544} 
\author{and Ue-Li Pen}
\affil{Harvard-Smithsonian Center
for Astrophysics, 60 Garden St., Cambridge, MA 02138}
\newcommand{\etal}{{\it et al. }}

\begin{abstract}
The string-dominated universe locally resembles
an open universe, and fits dynamical measures of power spectra,
cluster abundances, redshift distortions, lensing constraints,
luminosity and angular diameter distance relations and
microwave background observations.  We show examples of networks which
might give rise to recent string-domination without requiring any
fine-tuned parameters.  We discuss how future observations
can distinguish this model from other cosmologies.
\end{abstract}

\section{Introduction}

Most theoretical cosmologists prefer flat universe models. While this
preference was initially based on extensions of the Copernican principle
(Dicke 1970), it has been strengthened by the theoretical successes of the
inflationary universe paradigm (see Linde 1990 for discussion). While it is
possible to construct inflationary models with $\Omega <1$ (e.g., Linde and
Mezhlumian 1994), these models are less aesthetically appealing than the
flat universe models.

Observations, however, suggest that the matter density of the universe
is not sufficient to make $\Omega =1:$ measurements of the Hubble
constant (Freedman, Madore \& Kennicutt 1997) and estimates of the age
of the universe (Bolte and Hogan 1995) suggest that $H_0t_0>2/3;$
measurements of the baryon to dark matter ratio in clusters, together
with estimates of the baryon density from big bang nucleosynthesis
imply $\Omega _0,$ the energy density in matter, is much less than
$1$ (White \etal 1993); and the power spectrum of large scale
structure is best fit by 
models with $\Omega _0h_0=0.25$ (Peacock \& Dodds 1994). Here, $%
h_0=H_0/(100$ km/s/Mpc).  For several decades, it has been observed
that the Mass-to-Light ratio in clusters of galaxies suggests
$\Omega_0 \sim  0.2$ (Bahcall, Lubin \& Dorman 1995).  The simplest
COBE normalized parameter-free Harrison-Zeldovich-Peebles power
spectrum slope $n=1$ 
predicts local peculiar velocities and cluster abundances in
$\Omega_0=1$ and $\Lambda$ universes which are significantly higher
than observed (Strauss and Willick 1995, Eke, Cole and Frenk 1996, Pen
1996a, Viana and Liddle 1995),
which can be resolved by lowering the matter density $\Omega_0$.

This contradiction has led cosmologists to consider exotic equations of
state for the universe. The most studied modification of the standard matter
dominated cosmology is the vacuum dominated model. While there is no
particle physics motivation for positing a vacuum energy of 10$%
^{-124}M_{Planck}^4$ (Weinberg 1996), the model does appear to be
consistent with a number of observations (Ostriker and Steinhardt 1995).
However, recent measurements of $q_0$ using distant supernova
(Perlmutter \etal 1996) and limits
based on the statistics of gravitational lensing (Kochanek 1996) are
encouraging cosmologists to consider alternative models.  A novel
technique of distance determination using cluster hydrostatic
equilibrium measurements also indicates positive values of $q_0$ (Pen
1996b). 

A string dominated cosmology is an intriguing alternative to the standard
model. In this model, the energy density in strings scales with the
expansion factor, $a,$ as $a^{-2}$, decaying faster than a vacuum energy
term, but slower than the energy density in matter (which decays as $a^{-3}).
$ In this model, strings form at near the electroweak symmetry breaking
scale. Unlike the much heavier GUT scale strings (see e.g., Vilenkin and
Shellard 1993), these light strings do not seed structure formation.
Individual strings in this model have too low a density to be observable
separably. A typical string density would be $10^{-5}$ kg/m.
However, their culmative effect is to alter the expansion of the
universe. Locally, they make a flat universe appear to have many of the
properties of an open universe model.  The energy density of
such a string network arises naturally to be near the critical energy
density today.

If the universe today is string-dominated,
then the strings must be produced near the electroweak scale, a scale
at which there must be new physics.  These electroweak
strings are very light and would be undetected through
their gravitational lensing as their bending angle
is only $(M/M_{Pl})^2 \sim  10^{-32}$ radians.  Here,
$M$ is the symmetry breaking scale associated with
string formation and $M_{Pl}$ is the Planck scale.  While these
strings are light, they are expected to be numerous.  The
characteristic separation
between strings is expected to be the bubble size during the phase
transition, which in the case of the electro-weak phase transition is
typically $10^{-3}$ of the horizon size (Moore and Prokopec 1996), 0.1 A.U.(comoving). Thus, there would be many light strings in
our own Solar System.  If these light strings are associated
with baryogenesis (Starkman \& Vachaspati 1996) or are superconducting
(Vilenkin 1989),
then they may be directly detectable.

Only some cosmic string models lead to a string dominated universe. In
theories where cosmic strings can intercommute, their evolution obeys a
``scaling solution'': their energy density scales as $a^{-3}$ during matter
domination and as $a^{-4}$ during radiation domination. In these theories,
strings never dominate the energy density of the universe. \ On the other
hand, if strings do not intercommute nor pass through each other, then
the network can ``freeze-out'' 
and the energy density in strings can scale as $a^{-2}$ (Vilenkin
1984). Initial interest in 
string dominated universes was spurred by the possibility that Abelian
strings might not intercommute effectively and might dominate the energy
density of the universe (Kibble 1976; Vilenkin 1984; Kardashev 1986).
However, numerical simulations showed that even complicated Abelian string
networks (Vachaspati and Vilenkin 1987) intercommuted effectively and
rapidly approached the scaling solution. Despite the lack of a model that
had non-intercommuting strings, the interesting astrophysical implications
of a string dominated universe led to a number of papers investigating their
cosmological properties (Turner 1985; Charlton and Turner 1987; Gott and
Rees 1987; Dabrowski
and Stelmach 1989; Tomita and Watanabe 1990; Stelmach, Dabrowski and Byrka
1994; Stelmach 1995) and the properties of  cosmologies
with similar equations of state (Steinhardt 1996; Coble, Dodelson
\& Frieman 1996).   We review some of these results in section 3 and
compare string-dominated flat cosmologies to observations of large-scale
structure, microwave background fluctuations, observations of rich clusters,
and other cosmogical probes. In this section, we show that a
string-dominated universe with $H_0\sim 60-65$ km/s and $\Omega _0%
\sim 0.4- 0.6$ agrees remarkably well with a broad class of
observations.

Our interest in string-dominated universes was stimulated by our numerical
simulations of the evolution of Non-Abelian cosmic strings (Pen and Spergel
1996). In theories in which a non-Abelian symmetry is broken to a
discrete sub-group, multiple types of cosmic strings can be produced (Mermin
1979). Topological constraints prevents these different types of strings
from intercommuting (Toulouse 1976; Poenaru and Toulouse 1977), which
led to
the speculation that they could potentially dominate the energy density of
the universe (Kibble 1980). There are a number of phenomenologically
interesting particle physics models that utilize these non-Abelian
symmetries (Chkareuli 1991; Dvali \&\ Senjanovic 1994). These complex
string networks are not merely flights of theoretical fantasy: they can be
seen in biaxial nematic liquid crystals ( De'Neve, Kleman and Navard 1992).
In section 2, we discuss the physics of non-Abelian strings and summarize
the results of our numerical simulations.

\section{Physics of Non-Abelian Strings}

\subsection{What are Non-Abelian Strings?}

Strings are created when the lowest energy state of an order
parameter is degenerate and its vacuum manifold not simply connected.
A simple such 
example is given by nematic liquid crystals.  Each crystal is a needle
with perfect reflection symmetry.  The unbroken symmetry state above
the liquid crystal phase transition is one where each molecule can
point in any random direction in space, which is described by the
rotation group $G=SO(3)$.  In the liquid crystal phase, neighboring
elements prefer to point in the same direction, but rotation around
the needle axis is not distinguishable, nor are reflections around the
the plane perpendicular to its axies.  The broken symmetry group is
$H=O(2)$.  The vacuum 
manifold is given by ${\cal M}=G/H$, and its first homotopy group satisfies
the exact sequence 
\begin{equation}
\pi_1(H) \rightarrow \pi_1(G) \rightarrow
\pi_1(G/H) \rightarrow \pi_0(H) \rightarrow \pi_0(G).
\label{eqn:homot}
\end{equation}
Since
$\pi_0(SO(3))=0$ and $\pi_0(H)=Z_2$, we know that $\pi_1(G/H)$ must be
non-trivial.  In fact, $\cal M$ is just the projective 2-sphere, the
unit sphere with antipodes identified.  We know that $\pi_1({\cal M}) =
Z_2$, which is an Abelian group with only two elements, one of which
is the identity element.  All strings correspond to the other element,
allowing any 
two strings to intercommute.

The situation gets more interesting when different types of strings
can be formed.  Two strings corresponding to distinct group elements
of $\pi_1$ can intercommute, i.e. exchange partners, only if they
correspond to the same element, or to the inverses of each other.
When that is not the case, they can still pass through each other if
their corresponding elements commute.  In a non-Abelian system
there exist elements which do not commute, and two strings which
attempt to cross each other result in a configuration where an
umbilical cord is formed between the points where they crossed.

In general, each string type may have a different tension $\mu$, and
strings can decay into factors if that is energetically favorable.
If the tension in the umbilical cord is larger than twice the tension
in either of the intersection strings, it is energetically favorable
to have an umbilical cord of zero length, which macroscopically appear
like the junction of four string segments at one vertex.  Similarly, a
vertex joining any number of strings may be formed, depending on the
exact structure of $\pi_1$ and the distribution of string tensions.

\subsection{Dynamics of a Biaxial Nematic Crystal}

A particularly illustrative example of a system exhibiting non-Abelian
string defects are the biaxial nematic crystals.  Each crystal element
is in this case triaxial, with symmetry for 180 degree rotation
about any of the three axes.  This broken symmetry state is described
by the four element dihedral group $H=D_4$, one element corresponding
to no rotation, and the other three to a 180 degree rotation about
each coordinate axis.  Such a system is available commercially, and
has been studied experimentally by
Zapotocky, Goldbart and Goldenfeld (1995).

The exact homotopy sequence (\ref{eqn:homot}) describes the system.
We have $0\rightarrow 0\rightarrow Z_2\rightarrow \pi_1({\cal M})
\rightarrow D_4 \rightarrow 0$.  We thus know that $\pi_1({\cal M})$
must be an 8 element group, which in this case is the quaternion group
$Q_8$ with 8 elements $(1,i,j,k,-1,-i,-j,-k)$ and the multiplication
properties $i^2=j^2=k^2=-1,\ ij=k,\ jk=i,\ ki=j$.  The non-Abelian
property is exhibited by $ij=-ji$ etc, which derives from the
commutation property of the rotation group.  We have seven different
strings in this system, which can have up to four different tensions.
In a liquid crystal system, the string corresponding to $-1$ can
always decay into two strings from the generator with the smallest
tension.  The only non-commuting 
strings are those corresponding to $i,j,k$, and when they cross, they
in principle create an umbilical cord with charge $-1$, but
energetically they prefer to stick, forming a four leg vertex.  Three
leg vertices form at the junction of $i,j,k$ strings (or their
inverses).

If we neglect the presence of sticking strings (four leg vertices),
the system is quite similar to the $Z_3$ monopole-string network
studied by Vilenkin and Vachaspati (1987).  Whenever two three-leg
vertices come together, they annihilate and release two disconnected
strings.  To simulate these and other networks more realistically, we have
developed a global string code which simulates non-Abelian strings
using a nonlinear $\sigma$ model on a lattice.

\subsection{Dynamics of More Complex String Networks}

To simplify the simulation while capturing the essentials of a wide
range of non-Abelian string dynamics, we chose a modification of the
non-linear $\sigma$ model from Pen, Spergel and Turok (1995)
(hereafter PST).  In this
model, we have a classical field $\phi$ defined at every lattice point
$\vec{x}$, which takes on values in the range $[0,\pi)\times Z_N$.  The
field has both a continuous component, and a discrete index in the
range $0..N-1$.  The multiple leaves of semicircles are to be thought
of as a rolodex filer:  Whenever we examine the dynamics of two
leaves, we open the system such that the two leaves form a full
circle, and treat the dynamics as lying in a unit circle.  Since the
evolution equation only require the pairwise force, any two points
always lie on some such circle.

\begin{figure}
\centerline{\epsfxsize=\hsize \epsfbox{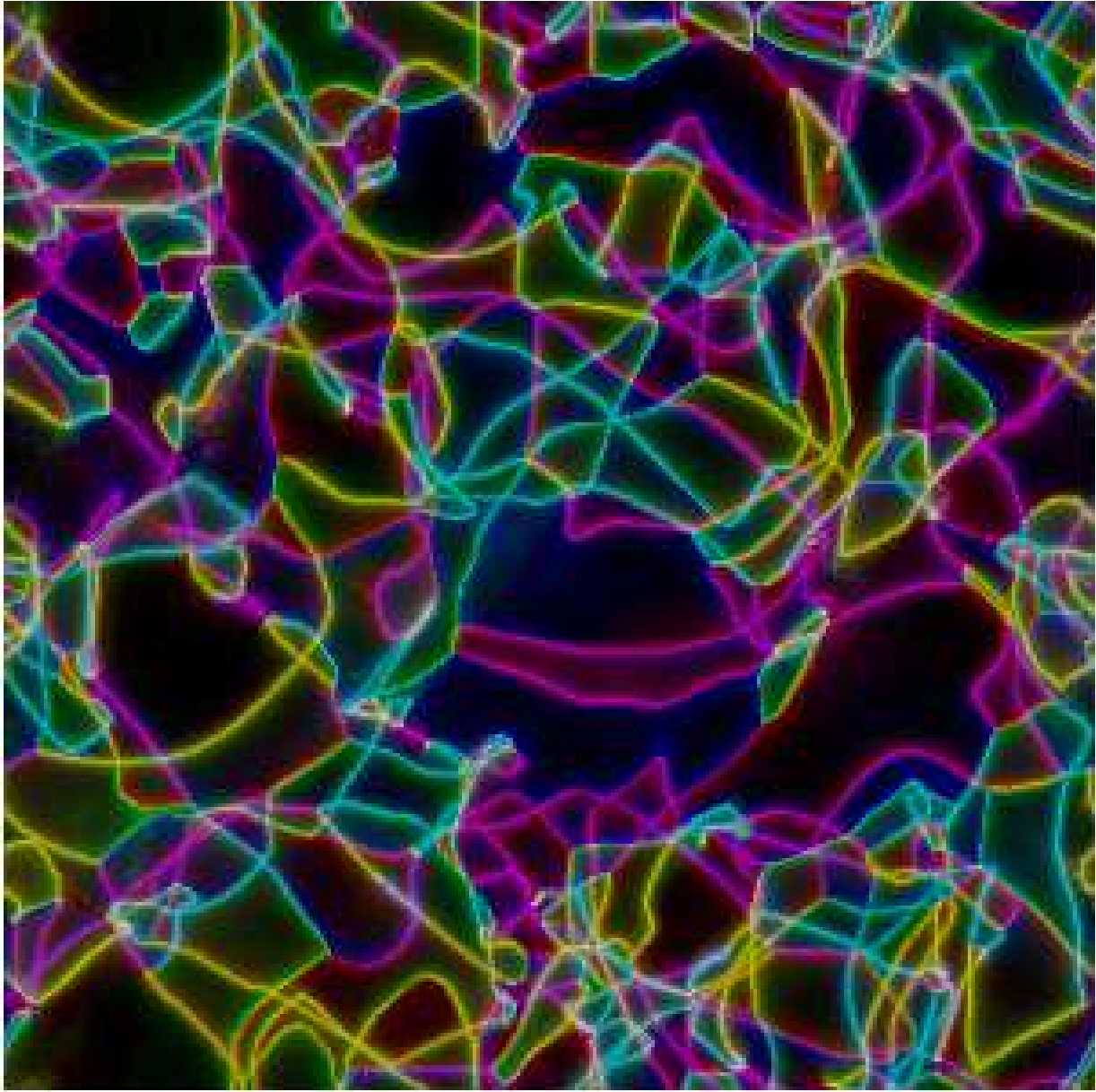} }
\caption{The network of a $N=3$ string system, which exhibits dynamics
very similar to biaxial nematic liquid crystals.  The strings are
color coded according to the generator they belong to.} 
\label{fig:network}
\end{figure}

For $N=2$, we recover the standard global strings from PST.  For
$N=3$, a system very simular to the biaxial nematics and the $Z_3$
strings is obtained, with three different types of strings, and three
point vertices which annihilate pairwise.  Two strings get stuck
when they try to pass through each other, just like the biaxial
nematic liquid crystals.  We show such a network in figure
\ref{fig:network}, where we have represented the string corresponding
to each of the three generators by a different color.  Each of the
three semicircle leaves are either red, green or
blue.  Since each string contains a complete rotation which covers two
leaves, the strings appear as composite colors, green+red=yellow,
etc.  There are three such possible pairs.

In general, we have a system of $N(N-1)/2$
strings.  Strings join at three-point vertices, of which there are
$N(N-1)(N-2)/6$ different types.  When two vertices join, there is a
one in $3/N$ chance that they can annihilate and
result in two disconnected strings.  Otherwise, the two vertices can
pass through one another and result in a new configuration which still
contains the same number of vertices.

\begin{figure}
\centerline{\epsfxsize=\hsize \epsfbox{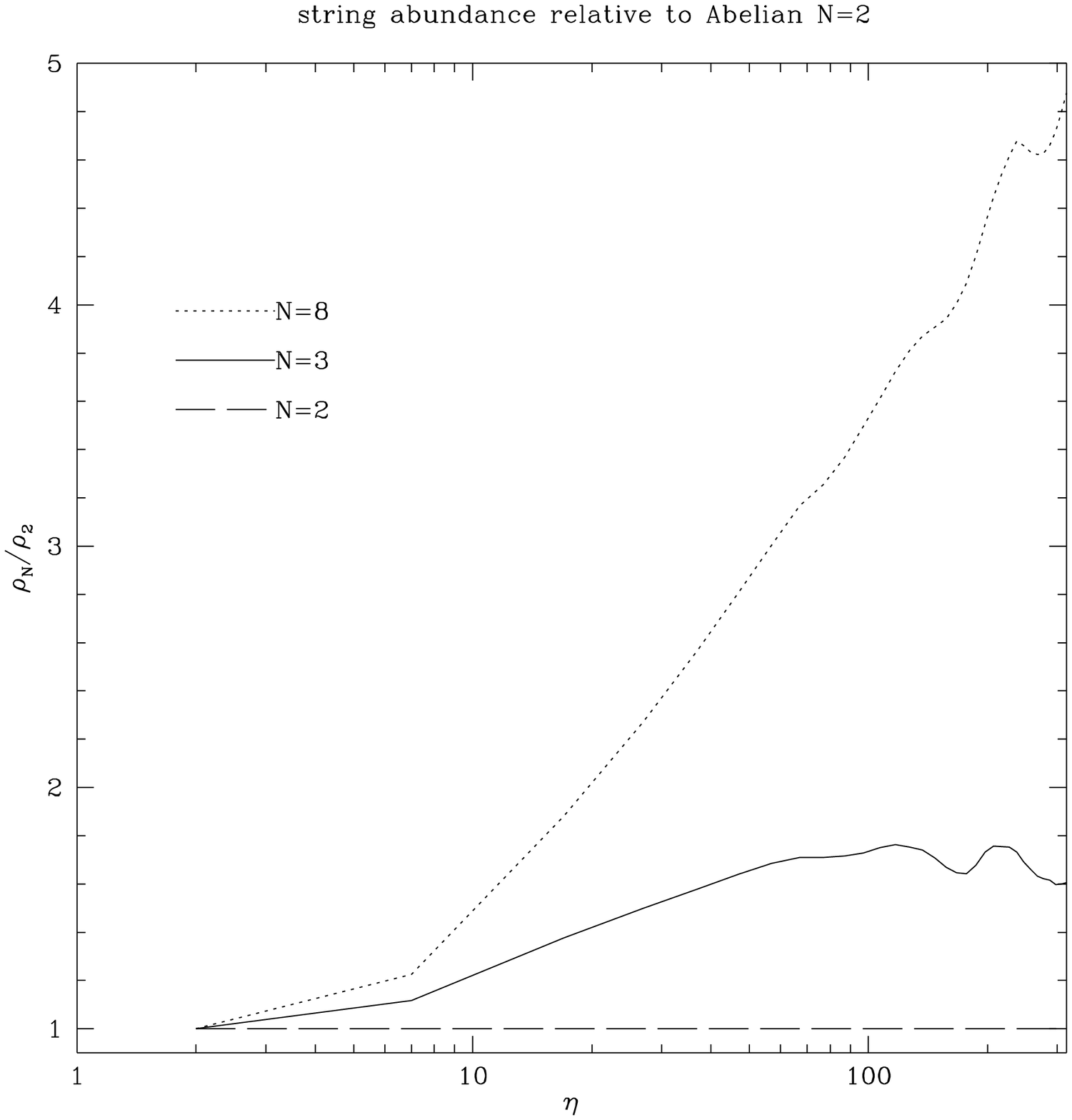} }
\caption{This figure shows the evolution of the string density (normalised
to the scaling density) as a function of time
in different models.  In scaling solutions, the string density should asymptote
to a constant value in this plot.  Note that the large $N$ models
do not scale and become tangled.}
\label{fig:corrn}
\end{figure}

We can now vary the number of string generators $N$ to correspond to a
one parameter class of non-Abelian strings.  We expect strings to
become more strongly tangled as we increase $N$.  This is indeed
observed, as shown in figure \ref{fig:corrn}.  The global field
dynamics differs systematically from gauged strings in the fact that
global strings exert long-range forces on each other, which can cause
the network to move even when the configuration is neutrally stable.
A neutrally stable solution, such as a sheet of hexagonal tilings, is
sufficient to cause full entanglement for cosmological purposes, since
the damping due to the universal expansion would prevent the structure
from collapsing.

We have found that for $N=3$ strings, the solution scales much like
the $Z_3$ monopole-string network, which would suggest that the
biaxial nematic liquid crystal system would also exhibit scaling
behaviour  (Pen and Spergel 1996).  We also see that the network does
seem to stop disentangling for large $N$.  While no
current model of the electro-weak phase transition predicts cosmic
strings, electro-weak baryogenesis calculations have argued for the
presence of more complicated symmetry structures.  It would be
conceivable that both the baryon asymmetry and the present day vacuum
energy be caused by the electro-weak symmetry breaking, which may have
testable laboratory consequences in the near future.

\section{Astrophysics of String Dominated Universe}

\subsection{Expansion of the universe: $H_0,q_0,\Omega _0$ and $t_0$}

Non-commuting strings formed at low energies have only one basic effect on
the universe: they add an additional term to the Friedman equation that
governs the expansion of the universe:
\begin{equation}
H^2=\left( \frac{8\pi G}3\right) \left[ \rho _{s0}\frac{a_0}{a}^2+\rho
_{m0}\frac{a_0}{a}^3+\rho _{r0}\frac{a_0}{a}^4\right]
\label{Hubble}
\end{equation}
where $H$ is the Hubble rate, $a$ the expansion factor and $\rho _{m0},\rho
_{r0},$ and $\rho _{s0}$ is the current energy density in matter, radiation
and strings. Since this additional term has the same $a$ dependance as the
presence of space curvature, a string-dominated flat universe is
observationally similar to matter-dominated open universe. Since we are
focusing on a flat universe, we define $\Omega _0\equiv 8\pi G\rho
_{m0}/3H_0^2,$ $\Omega _s\equiv 8\pi G\rho _{s0}/3H_0^2,$ $\Omega _r\equiv
8\pi G\rho _{r0}/3H_0^2,$ and assume $\Omega _0+\Omega _r+\Omega _s=1$

As in a curvature dominated universe, we can divide the history of the
universe into three epochs: a radiation dominated epoch, a matter dominated
epoch and a string-dominated epoch. In the matter and string dominated
epochs, we can express the evolution of the universe in terms of the
conformal time, $\eta \equiv \eta_*\cosh^{-1}(2\Omega^{-1}-1)$ (Peebles 1993):
\begin{eqnarray}
a &=&\frac{\Omega_0}{2(1-\Omega_0)}\left[ \cosh \eta/\eta_* -1\right]
\label{expansion} \\ 
H_0t &=&\frac{\Omega_0}{2(1-\Omega_0)^{3/2}}\left[ \sinh\eta/\eta_*-{\eta \over
\eta_* } \right] 
\nonumber
\end{eqnarray}
where we have defined $a_0=1$  and $\eta_*^{-1} = H_0 \sqrt{1-\Omega_0}$.

Thus, the relationship between the age of the universe, $t_0,$ the energy
density in matter, and the Hubble constant is the same as in a curvature
dominated universe:
\begin{equation}
H_0t_0=\frac{\Omega_0}{2(1-\Omega _0)^{3/2}}\left[ \frac{2}{\Omega_0}
(1-\Omega_0)^{1/2}-\cosh^{-1}(2\Omega_0^{-1}-1)\right]
\end{equation}
(Kolb and Turner 1990). We also recover the familiar
relationship for the deceleration parameter:
\begin{equation}
q_0=\frac{\Omega _0}2
\end{equation}
Thus, the string dominated cosmology makes the same predictions for most of
the classical cosmological tests as the open universe model.

Because the curvature of the string-dominated universe is flat, its angular
diameter-redshift relationship differs from an open universe. In an open
universe, the angular size  distance out to a redshift $z_e$ is 

\begin{equation}
H_0r(z_e)=\frac 1{\sqrt{1-\Omega _0}}\sinh \chi 
\end{equation}
(Peebles 1993), while in a string dominated flat universe,
\begin{equation}
H_0r(z_e)\;=\frac \chi {\sqrt{1-\Omega _0}}  \label{angular}
\end{equation}
where 
\begin{equation}
\chi =\int_{a_e}^1\frac{da}{\sqrt{\frac{\Omega _0}{1-\Omega _0}a+a^2}}.
\end{equation}
This altered angular diameter distance affects  number count predictions,
the probability of gravitational lensing and the predictions for microwave
background fluctuations. 
For both models, the number count statistics can be
computed from equation (13.61) in Peebles (1993):

\begin{equation}
\frac{d{\cal N}}{dz}=\frac{n_0}{H_0}\frac{r(z_e)^2}{\sqrt{\Omega
_0(1+z_e)^{-3}+(1-\Omega _0)(1+z_e)^{-2}}}
\end{equation}
where $n_0$ is the comoving density of galaxies. The string-dominated flat
model predicts fewer galaxy counts per unit redshift than both the open
universe model and the vacuum energy dominated model.

The statistics of gravitational lensing in this model differs significantly
from the predicted statistics in a vacuum dominated model. Current
observations already place strong constraints on the vacuum dominated model,
which predicts too many small lens events, particularly with small angular
separation (Turner 1990; Kochanek 1996). The absence of large number of
lenses in the HST\ snapshot survey (Maoz \etal 1993) and in radio surveys
implies that $\Omega _\Lambda <\;0.6$ and rules out most of the interesting
parameter space for vacuum dominated models. Because of the very different
relation between redshift and distance in string dominated models, it
predicts many fewer gravitational lenses than the vacuum dominated models. A
recent analysis by Bloomfield-Torres \&\ Waga (1995) finds that
string-dominated flat universes are excellent fits to the observed lens
statistics in the HST snapshot survey.

Observations of supernova at high redshift are another powerful
probe of cosmology.  Perlmutter \etal (1996) have already been able to
rule out cosmological constant models with $\Omega_0 < 0.6$ at the 95\%
confidence level with their supernova data.
Thus, there are no cosmological constant models compatible with
this observation, measurements of large scale structure,
measurements of the Hubble constant and the constraint that
the age of the universe exceed 11 Gyr.  At the redshifts
probed by the supernova study, the distance redshift relation
in a string-dominated universe is close to, but not identical
to the distance-redshift relation in an open universe.  Using
the relations given in equation (\ref{angular}), the
Perlmutter \etal (1996) observations imply that
$\Omega_0 > 0.15$ in a flat string-dominated cosmology.

%

\subsection{Microwave Background Fluctuations}

Because the strings only make significant contributions to the energy
density of the universe at very late times, they have no effect on the
physics at the surface of last scatter. However, since the strings alter the
expansion rate of the universe, they have two effects on the detailed shape
of the microwave background spectrum: (1) the decay of potential
fluctuations at late times produces additional fluctuations on large angular
scales; and (2) since the conformal distance to the surface of last scatter
is smaller, the Doppler peaks are shifted to larger angular scales
(Stelmach, Dabrowski, and Byrka 1994). Since identical effects occur in a
vacuum dominated universe, it will be difficult to
distinguish a string dominated universe from a vacuum dominated universe
based on CMB observations. On the other hand, it will be very easy to
distinguish a flat cosmology from an open universe due to the large
differences in the angular diameter distance relation (
Kamionkowski \& Spergel 1994).

\begin{figure}
\centerline{\epsfxsize=\hsize \epsfbox{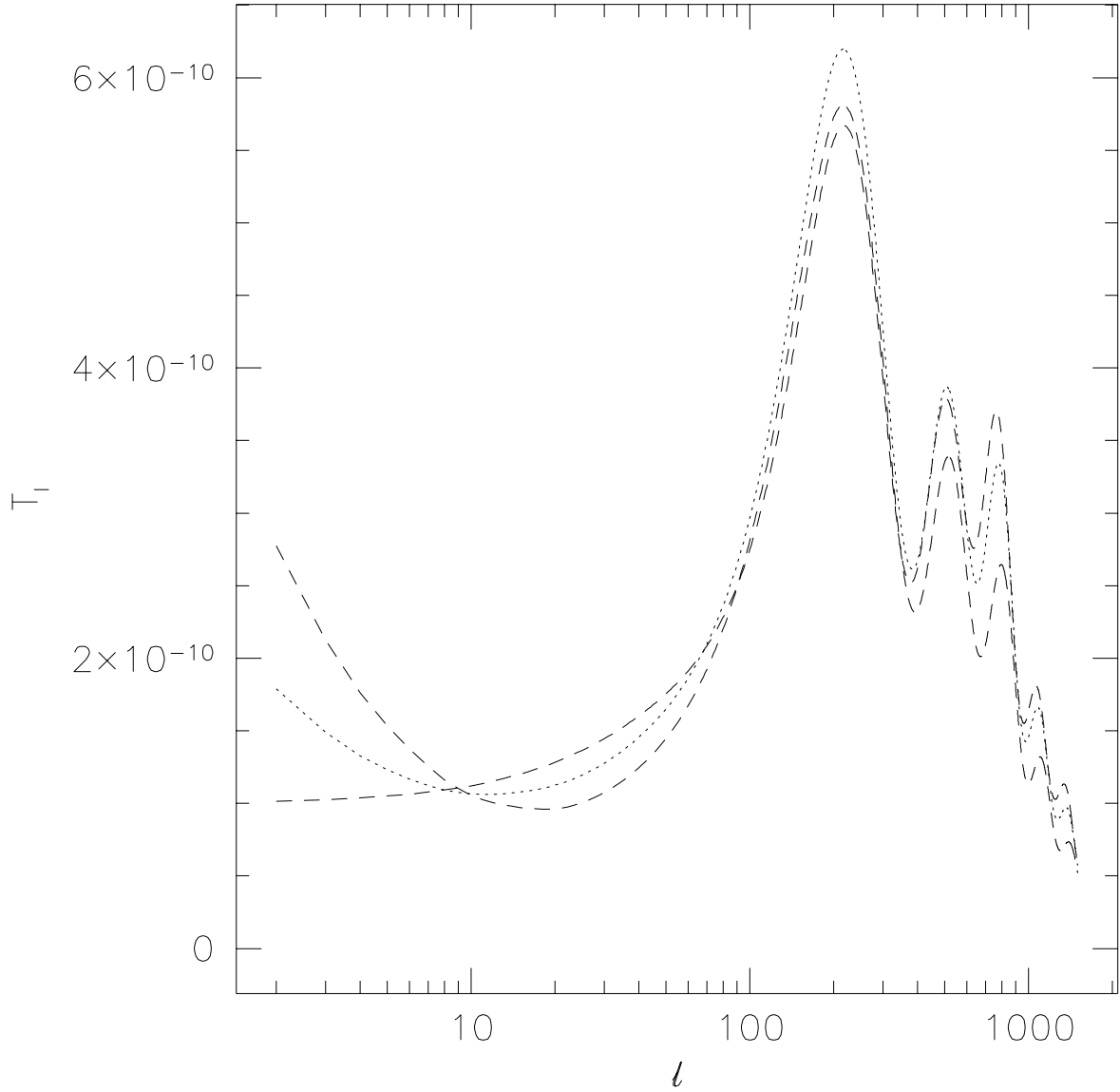} }
\caption{This figure compares the predicted multipole spectrum 
for three different models: a flat standard CDM model with $\Omega_0=1.0$
and $H_0 = 50$ km/s/Mpc (solid line);  a string-dominated flat cosmology
with $\Omega_0 = 0.4$ and $\Omega_0 = 0.6$.  Because COBE did not
detect a large quadrupole, the relative likelihood of the $\Omega_0 = 0.4$
to the $\Omega_0 = 1.0$ model is 0.05.}
\label{fig:cmb}
\end{figure}
\begin{figure}
\centerline{\epsfxsize=\hsize \epsfbox{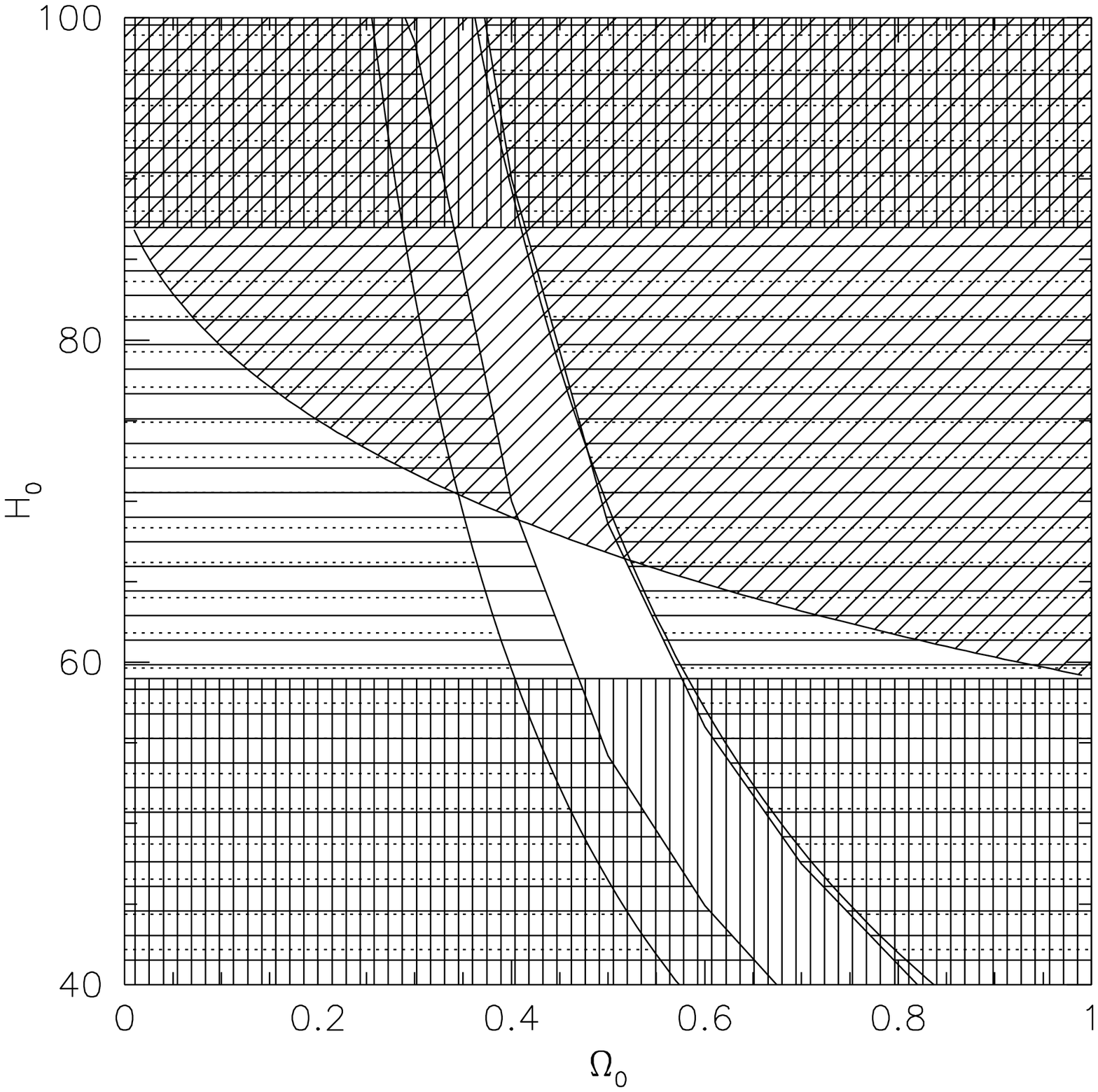} }
\caption{This figure combines constraints from various
astrophysical measurements.  The vertically shaded region
lie outside the best determinations of the Hubble
Constant: $H_0 = 73 \pm 6 \pm 8$  km/s/Mpc (Freedman, Madore \& Kennicutt
1997); the horizontally shaded regions do not agree with
measurements of the shape of the galaxy power spectrum,
$\Gamma = 0.25\pm 0.05$ (Peacock \& Dodds 1994), and with 
measurements of the fluctuation amplitude from clusters,
 $\sigma _8\Omega _0^{0.6} =0.6 \pm 0.1$ (Eke \etal 1996; Viana and
Liddle 1996; Pen 1996a); and the region shaded with lines
at 45$^o$ angle corresponds to cosmic ages less than 11 Gyr.}
\label{constraints}
\end{figure}

We have calculated the predicted CMB spectrum in a string-dominated
universe using a modified version of a Boltzmann code developed by Seljak and
Zaldarriaga (1996). Figure \ref{fig:cmb} shows the predicted multipole
spectrum for 
various string dominated cosmologies.  While
the three spectra can not yet be distinguished by
current observations, future CMB maps should be able easily
distinguish between the curves in figure \ref{fig:cmb}.

Most observations of large-scale structure are effectively measurements
of the galaxy power spectrum.  Peacock \& Dodds (1994) have shown
that most galaxy surveys are consistent with a standard
CDM power spectrum with $\Gamma \equiv \Omega_0 h \exp(-\Omega_b-\Omega_b/
\Omega_0) = 0.25 \pm 0.05 $.   The Las Campanas redshift survey
is also compatible with a power spectrum with $\Gamma = 0.2 - 0.3$
(Lin \etal 1996).  Figure \ref{constraints} shows that this
constraint alone is sufficient to rule out much of parameter space.
Note that standard CDM in a matter dominated flat universe
is ruled out unless $H_0 \sim  30$ km/s/Mpc.

We use the CMB spectrum to normalise the standard inflationary
model (scale-invariant, $\Omega_b h^2 = 0.0125$) in this cosmology to the
COBE observations.  Once this normalization is fixed,
there is no free parameters left in the model, so that
it can be compared directly to observations of matter power spectrum.

Observations of clusters are powerful probes of the matter power spectrum.
Gravitational lensing observations, X-ray observations of hot gas,
and studies of galaxy kinematics in clusters, all probe the
velocity distribution in clusters.  Thus, they can constrain the distribution
of mass, rather than the distribution of light. A number of studies
(Eke \etal 1996, Viana \& Liddle 1996; Pen 1996a) have concluded
that these cluster observations place very strong constraints
on the amplitude of mass fluctuations on the 8$h^{-1}$ Mpc
scale: $\sigma_8 = 0.6 \pm 0.1 \Omega_0^{-0.6}$.  Figure
\ref{constraints} shows that most string-dominated models
fit all the constraints.

\section{Conclusions}

String-dominated cosmologies have a number of very attractive
features. 

For $\Omega_0 \sim  0.4 - 0.6$ and $H_0 \sim  60 - 70$ km/s/Mpc, 
the model is consistent
with current observations.  It fits observations of the CMB,
measurements of the shape of galaxy power spectrum and
measurements of the amplitude of the mass power spectrum,
and is compatible with age limits.  Unlike
cosmological constant models, string dominated cosmology is also
consistent with observations of high redshift supernova and
gravitational lensing statistics.

Unlike cosmological constant models, which require new physics
at a very low energy scale, $\sim  10^{-4}$ eV,  the string dominated
model requires the introduction of new physics at the TeV scale,
where unitarity arguments in the standard model {\it require}
new physics (Wilzcek 1996).

The observational predictions of the string dominated model are intermediate
between the open universe model and the vacuum energy (cosmological
constant) model. CMB\ observations can easily distinguish between an open
universe model and the flat universe models (string-dominated,
matter-dominated, or vacuum energy-dominated): the open universe model
predicts that the Doppler peak should occur at $l\sim 220\Omega
^{-1/2}.$ Low redshift measurements can distinguish between the matter,
vacuum energy and string dominated models: the string dominated model
predicts $q_0=\Omega _0/2,$ while the vacuum dominated model predicts $%
q_0=3\Omega _0/2-1.$ Thus, future observations should be able to determine
the equation of state of the universe.


\end{document}